\documentclass[fleqn,usenatbib]{mnras}

\usepackage{newtxtext,newtxmath}
 \usepackage{framed}
\usepackage{xcolor}
\usepackage{ulem}
\usepackage{soul}

\DeclareRobustCommand{\VAN}[3]{#2}
\let\VANthebibliography\thebibliography
\def\thebibliography{\DeclareRobustCommand{\VAN}[3]{##3}\VANthebibliography}

\usepackage{graphicx}	
\usepackage{amsmath}	

\usepackage[T1]{fontenc}
\usepackage{verbatim}

\newcommand{\Gyr}             {\,{\rm Gyr}}

\title[multi-band gravitational wave background]{Exploring the multi-band gravitational wave background with a semi-analytic galaxy formation model}

\author[Zhencheng. Li et al.]{
Zhencheng Li$^{1,2}$,
Zhen Jiang$^{3}$,
Xi-Long Fan$^{4}$\thanks{E-mail: xilong.fan@whu.edu.cn},
Yun Chen$^{1,2}$,
Liang Gao$^{1,2,5}$,
Qi Guo$^{1,2}$,
Shenghua Yu$^{6}$
\\
$^{1}$National Astronomical Observatories, Chinese Academy of Sciences, Beijing 100101, China\\
$^{2}$School of Astronomy and Space Sciences, University of Chinese Academy of Sciences, Beijing, 100049, China\\
$^{3}$Department of Astronomy, Tsinghua University, Beijing 100084, China\\
$^{4}$School of Physics and Technology, Wuhan University, Wuhan 430072, China\\
$^{5}$School of Physics and Microelectronics, Zhengzhou University, Zhengzhou 450001, China\\
$^{6}$CAS Key Laboratory of FAST, National Astronomical Observatories, Chinese Academy of Sciences, 20A Datun Road,Beijing 100101, China
}

\date{Accepted XXX. Received YYY; in original form ZZZ}

\pubyear{2022}

\begin{document}
\label{firstpage}
\pagerange{\pageref{firstpage}--\pageref{lastpage}}
\maketitle

\begin{abstract}
An enormous number of compact binary systems, spanning from stellar to supermassive levels, emit substantial gravitational waves during their final evolutionary stages, thereby creating a stochastic gravitational wave background (SGWB). We calculate the merger rates of stellar compact binaries and massive black hole binaries using a semi-analytic galaxy formation model -- Galaxy Assembly with Binary Evolution (GABE) in a unified and self-consistent approach, followed by an estimation of the multi-band SGWB contributed by those systems. We find that the amplitudes of the principal peaks of the SGWB energy density are within one order of magnitude $\Omega_{GW} \sim 10^{-9}- 10^{-8}$. This SGWB could easily be detected by the Square Kilometre Array (SKA), as well as planned interferometric detectors, such as the Einstein Telescope (ET) and the Laser Interferometer Space Antenna (LISA). The energy density of this background varies as $\Omega_{GW} \propto f^{2/3}$ in SKA band. The shape of the SGWB spectrum in the frequency range $\sim[10^{-4}$,$1]$Hz could allow the LISA to distinguish the black hole seed models. The amplitude of the SGWB from merging stellar binary black holes (BBHs) at $\sim 100$ Hz is approximately 10 and 100 times greater than those from merging binary neutron stars (BNSs) and neutron-star-black-hole (NSBH) mergers, respectively. Note that, since the cosmic star formation rate density predicted by GABE is somewhat lower than observational results by $\sim 0.2$ dex at z < $\sim 2$, the amplitude of the SGWB in the frequency range $\sim[1$, $10^{4}]$ Hz may be underestimated by a similar factor at most.
\end{abstract}




\section{Introduction} 

The present and forthcoming experiments aimed at detecting gravitational waves (GWs), including the advanced Laser Interferometer Gravitational wave Observatory (aLIGO) \citep{2010CQGra..27h4006H}, the Einstein Telescope (ET) \citep{2010CQGra..27s4002P}, the Laser Interferometer Space Antenna (LISA) \citep{2017arXiv170200786A}, and the Square Kilometre Array (SKA) \citep{2009IEEEP..97.1482D}, herald the advent of multi-band GW astronomy. Joint detection by multiple GW experiments holds the potential to realize the goal of exploring the multi-band stochastic GW background (SGWB), which arises from the energy background resulting from the superposition of various resolved and unresolved GW sources spanning a broad range of frequencies. These sources can be classified into two categories, namely, cosmological sources and astrophysical sources. The joint detection of the multi-band SGWB from cosmological sources has been extensively investigated \citep[e.g.][]{2021PhRvD.103l3541B,2021JCAP...01..012C}. Here, we concentrate on the joint detection of the multi-band SGWB originating from astrophysical sources.

Previous studies on the SGWB from astrophysical sources have primarily focused on specific types of GW sources (e.g., see recent reviews such as \citet{2019RPPh...82a6903C} and references therein). Among them, extensive investigations have been conducted on the SGWB generated by merging massive black hole (in this work, i.e., the central black hole of the galaxy) binaries \citep[e.g.][]{2004ApJ...611..623S, 2017MNRAS.464.3131K, 2018MNRAS.477.2599B,2019MNRAS.483..503Y, 2022MNRAS.509.3488I}. These investigations have typically estimated the evolution and merger rates of massive black hole binaries through cosmological hydrodynamic simulations or semi-analytic galaxy formation models. Additionally, numerous studies have explored the SGWB generated by merging stellar compact binaries, including stellar binary black holes (BBHs) \citep[e.g.][]{2011PhRvD..84l4037M, 2018PhRvD..98d4020F, 2019PhRvD.100f1101A, 2021PhRvD.104b2004A, 2021PhRvD.103l3541B, 2021MNRAS.500.1421Z, 2022A&A...660A..26B, 2022PhRvD.105j3032P}, binary neutron stars (BNSs) \citep[e.g.][]{2011ApJ...729...59Z, 2013MNRAS.431..882Z, 2018PhRvL.120i1101A, 2019ApJ...871...97C, 2022PhRvD.105b2001L}, neutron-star–black-hole binaries (NSBHs) \citep[e.g.][]{2018MNRAS.475.1331T, 2021JCAP...11..032C, 2021PhRvD.103d3002P}. Moreover, extreme mass-ratio inspirals, i.e., merging binary systems consisting of a compact stellar mass object orbiting a massive black hole, have also been extensively investigated \cite[e.g.][]{2020PhRvD.102j3023B,2023MNRAS.tmp..424W}.

In the near future, joint detection of the SGWB in different bands could provide an opportunity to assess the reliability of semi-analytic galaxy formation models and binary population synthesis models in a way that is independent of traditional electromagnetic observations. While detectors such as ET and LISA could directly detect loud GW events \citep[e.g.][]{2004ApJ...611..623S,2011PhRvD..84h4004R,2021PhRvD.103d3002P,2021JCAP...10..035B}, the shape and amplitude of the total SGWB provide additional information to constrain the population properties of source models. A previous study by \citet{2011PhRvD..84h4004R} investigated the multi-band SGWB from astrophysical sources using independently calculated merger rates of massive black hole binaries and stellar binaries. In this paper, we employ a unified and self-consistent approach to calculate the merger rates of compact binary systems ranging from stellar to supermassive levels. We derive the merger rates of stellar compact binaries and massive black hole binaries from a semi-analytic galaxy formation model called Galaxy Assembly with Binary Evolution (GABE) \citep{2019RAA....19..151J}, combined with a rapid binary population synthesis model  COSMIC v3.3.0 \citep{2020ApJ...898...71B}.
We comprehensively assess the influence of cosmic star formation rate density (SFRD) on the SGWB by incorporating empirical constraints for the SFRD. We also improve the completeness of the sample of intermediate-mass ($100 M_{\odot} < M < 10^6 M_{\odot}$) black holes by utilizing the Millennium-II simulation \citep{2009MNRAS.398.1150B} in place of the Millennium simulation \citep{2005Natur.435..629S} used in \cite{2019RAA....19..151J}.

The paper is organized as follows: firstly, the basic equations employed for the estimation of the SGWB signal are presented in Section~\ref{sec:Formalism}. Next, the methodology utilized for generating mock samples of GW sources is demonstrated in Section~\ref{sec:Simulation of sources}. In Section~\ref{sec:Results}, the SGWB signals are computed from different astrophysical sources which are expected to be distributed across a broad range of frequencies, followed by a presentation of the main findings. Finally, the primary conclusions drawn from the study are summarized in Section~\ref{sec:Summary and Discussion}.

\section{Formalism}
\label{sec:Formalism}

\subsection{Merger rates and SGWB}
\label{subsec:Merger rates and SGWB}

The merger rate of compact binary systems, which denotes the event number per comoving volume per cosmic time at redshift $z$, can be described as: 
\begin{equation}
    R(z) = \mathcal{N}(z)\frac{\mathrm{d} z}{\mathrm{d} t},
	\label{eq:merger rate}
\end{equation}
where $\mathcal{N}(z)$ is the number density (i.e., the number of GW sources in per comoving volume per unit redshift at redshift $z$).  
Throughout the paper, we adopt a fiducial cosmological model with
 $\Omega_\mathrm{m}=0.25$, $\Omega_\Lambda=0.75$ and $H_0=73\,\mathrm{km}/\mathrm{s}/\mathrm{Mpc}$ \citep[WMAP1,][]{2003ApJS..148..175S}, which is also  adopted in the Millennium simulation.
Then, one can get the relation between the redshift $z$ and the cosmic time $t$, i.e., $\mathrm{d}z/\mathrm{d}t =H_0 \left ( 1+z \right )[\Omega _\mathrm{m}\left (  1+z\right )^{3}+\Omega _\Lambda ]^{0.5}$.

The dimensionless energy density of the SGWB, $\Omega _\mathrm{GW}$, is defined as: 
\begin{equation}
    \Omega _\mathrm{GW}\left ( f \right )= \frac{1}{\rho _\mathrm{c}}\frac{\mathrm{d} \rho _\mathrm{GW}\left ( f \right )}{\mathrm{d} \mathrm{ln}\left ( f \right )},
	\label{eq:Omega origin}
\end{equation}
where ${\rho_\mathrm{c}}=3{{H_0}^2}{c^2}/8{\pi}G$ is the critical energy density, ${\rho_\mathrm{GW}}$ is the present-day energy density of GWs.  
According to \citet{2001astro.ph..8028P}, ${\Omega_\mathrm{GW}}$ is calculated as: 
\begin{equation}
    \Omega _\mathrm{GW}\left ( f \right )= \frac{1}{\rho_\mathrm{c}}\int \mathcal{N}\left ( z  \right )\frac{1}{1+z}\left [ \frac{\mathrm{d} E\left ( f_\mathrm{r} \right )}{\mathrm{d} \mathrm{ln}\left ( f_\mathrm{r} \right )} \right ]_{f_\mathrm{r}=\left ( 1+z \right )f}\,dz,
	\label{eq:Omega Phinney}
\end{equation}
where $\mathrm{d} E\left ( f_r \right )/\mathrm{d} ln\left ( f_r \right )$ is the energy spectrum of a single GW  source in logarithmic interval, and $f_\mathrm{r} = (1+z)f$ is the frequency in the cosmic rest frame of the GW source, $f$ is the observed frequency. 

Following \citet{2008PhRvD..77j4017A} and \citet{2011ApJ...739...86Z}, we adopt the GW energy spectrum with
\begin{equation}
\begin{aligned}
\frac{\mathrm{d} E(f_r, m_1, m_2)}{\mathrm{d} f_r}=& \frac{\left ( \pi G \right )^{2/3}M_\mathrm{c}^{5/3}}{3} \\ 
& \times
\left\{\begin{aligned}
&f_r ^{-1/3}, \, &f_r< f_{\mathrm{merg}} \\ 
&\omega_{1} f_r ^{2/3}, \,& f_{\mathrm{merg}}\leq f_r< f_{\mathrm{ring}} \\ 
&\omega_{2} \left [ \frac{f_r}{1+\left ( \frac{f_r-f_{\mathrm{ring}}}{\sigma /2} \right )^2} \right ]^{2}, \, & f_{\mathrm{ring}}\leq f_r< f_{\mathrm{cut}}
\end{aligned}\right.,
\end{aligned}
\label{eq:gw wave}
\end{equation}
where $M_\mathrm{c}=(m_1m_2)^{3/5}(m_1+m_2)^{-1/5}$ is the chirp mass, $m_1$ and $ m_2$ are the masses of the primary and secondary stars in the GW source; $\omega_1=f_{\mathrm{merg}}^{-1}$ and  $\omega_2=f_{\mathrm{merg}}^{-1}f_{\mathrm{ring}}^{-4/3}$ are constants to make spectrum continuous at the boundary; $f_{\mathrm{merg}}, f_{\mathrm{ring}}$ and  $f_{\mathrm{cut}}$ are characteristic frequencies to divide the different stages of the compact binaries. The parameters $f_{\mathrm{merg}}, f_{\mathrm{ring}}, \sigma$  and $f_{\mathrm{cut}}$ can be calculated with the formula $c^3(a_1\eta^2+a_2\eta+a_3)/\pi GM$ proposed in \citet{2008PhRvD..77j4017A}, where $\eta=m_1m_2/M^2$ is the symmetric mass ratio with $M=m_1+m_2$. The values of the coefficients [$a_1$, $a_2$, $a_3$] are [$2.9740 \times 10^{-1}$, $4.4810 \times 10^{-2}$, $9.5560 \times 10^{-2}$], [$5.9411 \times 10^{-1}$, $8.9794 \times 10^{-2}$, $1.9111 \times 10^{-1}$], [$5.0801 \times 10^{-1}$, $7.7515 \times 10^{-2}$, $2.2369 \times 10^{-2}$], and [$8.4845 \times 10^{-1}$, $1.2848 \times 10^{-1}$, $2.7299 \times 10^{-1}$] for $f_{\mathrm{merg}}, f_{\mathrm{ring}}, \sigma$ and $f_{\mathrm{cut}}$, respectively, which are also presented in the Table I of \citet{2008PhRvD..77j4017A}.  The Eq.~(\ref{eq:gw wave}) is developed to describe inspiral-merger-ringdown stages for  non-spinning merging black hole binaries on circular orbits, but we assume that for BNSs and NSBHs, the spectrum is still suitable, as in previous studies \citep[e.g.][]{2019ApJ...871...97C, 2021JCAP...11..032C}. 

Two other quantities are also used to describe the SGWB besides ${\Omega_\mathrm{GW}(f)}$: the one-sided spectral power density $S_\mathrm{h}(f)$, which is related to the detection criterion  (e.g. the signal to noise ratio) , and the characteristic strain $h_c(f)$. They are related to ${\Omega_\mathrm{GW}(f)}$ with the expression:
\begin{equation}
     fS_\mathrm{h}(f) = h_\mathrm{c}^{2}(f) = \frac{3H_0^{2} }{2\pi^{2} }f^{-2}\Omega _\mathrm{GW}(f),
	\label{eq:characteristic strain}
\end{equation}
The dimensionless characteristic strain $A_{\mathrm{yr}^{-1}}$ is usually introduced in the pulsar timing array (PTA) experiment, and the relation between $A_{\mathrm{yr}^{-1}}$ and  $h_\mathrm{c}(f)$ is
\begin{equation}
    h_\mathrm{c}(f) =A_{\mathrm{yr}^{-1}}\left(f / f_{\mathrm{yr}}\right)^{\alpha},
	\label{eq:characteristic strain power low}
\end{equation}
where the equality assumes a power-law form for the background and $f_{\mathrm{yr}}= 1 yr^{-1}$ is the reference frequency. Note that for the SGWB from supermassive ($M>10^6 M_{\odot}$) binary black holes, in the frequency $\sim 10^{-9} - 10^{-7}$ Hz, $\alpha \approx -2 / 3$ \citep[e.g.][]{2019MNRAS.482.2588Z}.

The merger rate $R(z)$ and the energy density of SGWB $\Omega_\mathrm{GW}(f)$ are both related to the number density $\mathcal{N}(z)$ as presented in Eqs.~(\ref{eq:merger rate}) and~(\ref{eq:Omega Phinney}).  
An analytical method to calculate $\mathcal{N}(z)$ (and then $R(z)$ or $\Omega_\mathrm{GW}(f)$) is carried out by using an analytical model for SFRD and metallicity evolution history \citep[e.g.][]{2020ApJ...899L...1Z} and combining  with the properties of merging compact binaries.
Alternatively, these quantities also can be computed by using the semi-analytic galaxy formation model. The later method is employed in this work as discussed in Section~\ref{sec:Simulation of sources}. In the semi-analytic method, one can compute $\mathcal{N}(z)$ with
\begin{equation}
    \mathcal{N}(z)=\frac{\Delta N(z)}{\Delta V_\mathrm{c}\Delta z},
\label{eq:N}
\end{equation}
where $\Delta N$ is the number of merging stellar compact binaries in the redshift interval $\Delta z$, and $\Delta V_c$ is the volume of the simulation box. 
Eq.~(\ref{eq:Omega Phinney}) can be rewritten as:
\begin{equation}
    \Omega _{\mathrm{GW}}\left ( f \right )= \frac{f}{\rho _\mathrm{c}}\sum^{\mathrm{num}}\frac{1}{\Delta V_\mathrm{c}}\left [ \frac{\mathrm{d} E\left ( f_\mathrm{r}, m_1, m_2 \right )}{\mathrm{d}  f_\mathrm{r}} \right ]_{f_\mathrm{r}=\left ( 1+z \right )f},
\label{eq:Omega in use directly}
\end{equation}
where '$\mathrm{num}$' is the total number of the GW sources. Furthermore, Eq.~(\ref{eq:Omega in use directly}) can also be calculated in terms of  $z$, $m_1$, and $m_2$  bins as:
\begin{equation}
\begin{aligned}
    \Omega _\mathrm{GW}\left ( f \right )= &\frac{f}{\rho _\mathrm{c}}\sum^{l}\sum^{m \times n}\frac{\Delta N(z, m_1, m_2)}{\Delta V_\mathrm{c}} \\
    & \times \left [ \frac{\mathrm{d} E\left ( f_\mathrm{r}, m_1, m_2 \right )}{\mathrm{d}  f_\mathrm{r}} \right ]_{f_\mathrm{r}=\left ( 1+z \right )f},
\end{aligned}
\label{eq:Omega in use}
\end{equation}
where $m$ and $n$ are the bin numbers of $m_1$ and $m_2$, respectively, which are determined by the division of the $m_1 \times m_2$ space, $l$ is the bin number of redshift $z$, which is determined by the division of redshift, $\Delta N$ is the number of merging stellar compact binaries in the redshift-primary mass-secondary mass interval $\Delta z, m_1, m_2$. In this work, we assume that the values of $\mathrm{d} E\left ( f_{r}, m_1, m_2 \right ) / \mathrm{d} f_{r}$ are equal to their values at the bin center.

\subsection{Signal to Noise ratio}
\label{subsec:Signal to Noise ratio}

We first describe the formalism we used to calculate the signal to noise ratio (SNR) for ground-based and space-based detectors. The expected SNR for a cross-correlation search for an unpolarized and isotropic SGWB is given by \citet{1999PhRvD..59j2001A, 2013PhRvD..88l4032T}. Once the SGWB is calculated, we can estimate the SNR for the given noise power spectral density $P_\mathrm{n}$ and overlap reduction function $\Gamma_{IJ}(f)$. $\Gamma_{IJ}(f)$ describes the reduction in sensitivity to the SGWB of two detectors labeled by $I$ and $J$ due to their separation and non-optimal orientations \citep{1993PhRvD..48.2389F} (when $I = J$, we define the transfer function $R_I(f) = \Gamma_{II}(f)$). 
 A normalized overlap reduction function:
\begin{equation}
    \gamma_{IJ}(f) = \frac{5}{ \sin ^2\delta } \, \Gamma_{IJ}(f),
\label{eq:gamma}
\end{equation}
is often used to ensure $\gamma_{IJ}(0) = 1$ for two identical, co-located and co-aligned detectors, where $\delta$ is the opening angle of the interferometer. In this work, we estimate the SNR with two detectors for both LIGO and ET. Assuming the two detectors are almost co-located, the $\gamma(f)\sin ^2\delta$ for LIGO can be approximated to $1$ \citep{2009PhRvD..79h2002N, 2008PhLB..663...17F} and for ET, it can be approximated to $-3/8$ \citep{2022PhRvD.105f4033A} over the frequency range we studied. For ground-based detectors,

\begin{equation}
    SNR = \frac{3H_0^2}{10\pi^2}\sqrt[]{2T}\left[\int_{f_\text{min}}^{f_\text{max}}df\frac{\gamma_{IJ}^{2}(f)\Omega_\mathrm{GW}^{2}(f)\sin ^4\delta}{f^6P_{\mathrm{n}I}(f)P_{\mathrm{n}J}(f)}\right]^{1/2},
    \label{eq:SNR ground}
\end{equation}
where $T$ is the observation period of the experiment, $f_\mathrm{max}$ and $f_\mathrm{min}$ are the upper and lower sensitivity bounds of the experiment. We adopt the $P_\mathrm{n}$ of aLIGO and ET from the LIGO document\footnote{https://dcc.ligo.org/LIGO-T1500293/public}. For LISA, the $\sqrt{2}$ in the Eq.~(\ref{eq:SNR ground}) can be reduced due to the use of data from only one detector \citep{2013PhRvD..88l4032T}:
\begin{equation}
    SNR = \sqrt[]{T}\left[\int_{f_\mathrm{min}}^{f_\mathrm{max}}df\frac{S_\mathrm{h}^{2}(f)}{S_\mathrm{n}^{2}(f)}\right]^{1/2},
    \label{eq:SNR sky}
\end{equation}
where $S_n(f) = P_n(f)/R_I(f)$ is the strain spectral sensitivity. We calculated $S_n(f)$ using the analytical formulas presented in \citet{2019CQGra..36j5011R}, considering the galactic confusion noise. 

For SKA, we use the python package $hasasia$ \citep{2019PhRvD.100j4028H} to calculate the SNR: 
\begin{equation}
    SNR \simeq \sqrt{2T} \left[\int_{0}^{f_{\text {Nyq}}} \mathrm{d} f \frac{S_{h}^{2}(f)}{S_{\mathrm{eff}}^{2}(f)}\right]^{1/2},
    \label{eq:SNR PTA}
\end{equation}
where $f_\mathrm{Nyq}=1/(2\Delta t)$, $\Delta t$ is the minimum time interval between two observations. $S_{\mathrm {eff }}(f)$ is the effective strain-noise power spectral density for the whole PTA in which a
single distinct pair is labeled by $I$ and $J$: 
\begin{equation}
    S_{\text {eff}}(f) \equiv\left(\sum_{I} \sum_{J>I} \frac{T_{IJ}}{T} \frac{\chi^2_{IJ}}{S_{I}(f)S_{J}(f)}\right)^{-1/2},
    \label{eq:S_eff}
\end{equation}
where $\chi_{IJ}$ are the Hellings and Downs factors, $S_I(f) = 1/(\mathcal{N}_{I}^{-1}(f) R_I(f))$ is the individual pulsar strain-noise power spectral density, $\mathcal{N}_{I}^{-1}(f)$ is the inverse-noise-weighted transmission function and $R_I(f) = 1/(12\pi^2f^2)$.

\subsection{Power-law integrated sensitivity curves}
\label{subsec:Power-law integrated curves}

Power-law integrated (PI) sensitivity curves are often used to estimate the detection abilities of various detectors, which takes into account the enhancement in sensitivity that comes from integrating over frequency \citep{2013PhRvD..88l4032T}. Assuming a set of power-law SGWBs $\Omega_{\beta}(f/f_{\mathrm{ref}})^{\beta}$ with indices e.g.,  $\beta=\{-8, -7, ..., 7, 8\}$ and an arbitrary reference frequency $f_{\mathrm{ref}}$, the PI sensitivity curves are the envelope of these SGWBs. 

By replacing $\Omega _{\mathrm{GW}}(f)$ with $\Omega_{\beta}(f/f_{\mathrm{ref}})^{\beta}$, one can calculate the amplitude $\Omega_{\beta}$ with Eqs.~(\ref{eq:SNR ground}), (\ref{eq:SNR sky}) and (\ref{eq:SNR PTA}) for ground-based detectors, LISA and SKA, respectively, where $S_h$ in Eqs.~(\ref{eq:SNR sky}) and (\ref{eq:SNR PTA}) is related to $\Omega _{\mathrm{GW}}(f)$ as displayed in Eq.~(\ref{eq:characteristic strain}).  
For example, for ground-based detectors, $\Omega_{\beta}$ is derived by replacing $\Omega_{\mathrm{GW}}$ with $\Omega_{\beta}(f/f_{\mathrm{ref}})^{\beta}$ in Eq.~(\ref{eq:SNR ground}):
\begin{equation}
    \Omega_{\beta} = \frac{10\pi^2\mathrm{SNR_{th}}}{3H_0^2\sqrt{2T}}\left[\int_{f_{\mathrm{min}}}^{f_{\mathrm{max}}}\mathrm{d}f \frac{(f/f_{\mathrm{ref}})^{2\beta}\gamma_{IJ}^{2}(f)\sin ^4\delta}{f^6P_{\mathrm{n}I}(f)P_{\mathrm{n}J}(f)}\right]^{-1/2},
    \label{eq:PI}
\end{equation}
where SNR$_{\mathrm{th}}$ is an arbitrary detection threshold.

\section{Simulation of sources}
\label{sec:Simulation of sources}

\subsection{Merging massive black hole binaries}
\label{subsec:Massive merging compact binaries}

Utilizing the Millennium (MSI) and Millennium-II (MSII) simulations in conjunction with a semi-analytic galaxy formation model--GABE, we simulate the emergence and growth of galaxies and their associated black holes in a cosmological context. While the MSII simulation can resolve all luminous galaxies and their associated black holes in the universe, it lacks sufficient volume to provide adequate statistics to model extremely large supermassive black holes, which can lead to noisy predictions of ultra-low frequency SGWB. To address this limitation, we leverage the MSI, which has a volume $125$ times greater than that of MSII, to model ultra-low frequency background.

As is common in many galaxy formation models, GABE assumes that a seed black hole is responsible for the formation of the central black hole of a galaxy. However, the exact mass of the seed black hole is not well-constrained by observations (see \citealt{2010A&ARv..18..279V,2019PASA...36...27W} for reviews). In this study, two different seed models are considered: the light-seed and heavy-seed models. The light-seed model proposes that the seed black holes are formed through the collapse of Population III stars \citep[e.g.][]{1984ApJ...280..825B,2001ApJ...551L..27M,2016MNRAS.458.3047P}, while the heavy-seed model suggests that they are the product of the direct collapse of atomic gas in dark matter halos with a virial temperature of about $10^4$K \citep[e.g.][]{2003ApJ...596...34B,2006MNRAS.370..289B,2006MNRAS.371.1813L,2017NatAs...1E..75R}. In practice, the light-seed model involves the addition of a $10^2 M_{\odot}$ black hole to a galaxy at emergence, while the heavy-seed model adds a $10^4 M_{\odot}$ black hole to a galaxy when it first appears. Notably, the choice of seed black hole mass has a negligible effect on subsequent stellar evolution or the final black hole mass in the semi-analytic models utilized in this study, as the seed mass only represents a small fraction of the final black hole mass ($\gtrsim10^6 M_{\odot}$). Additionally, any initial differences in seed mass are quickly erased by the following rapid growth during the first gas-rich galaxy merger. 

The growth model of massive black holes adopted in GABE is similar to that of \cite{2006MNRAS.365...11C}. In the early stages, seed black holes primarily grow through the process of merging with other black holes and accreting cold disk gas during galaxy mergers, which is commonly referred to as the 'quasar mode'. The growth during a merger event is:
\begin{equation}
    \delta M_\mathrm{BH}=M_\mathrm{BH,sat}+f\left ( \frac{M_\mathrm{sat}}{M_\mathrm{cen}} \right ) \left[\frac{M_\mathrm{cold}}{1+\left ( 280 \mathrm{km}/\mathrm{s}/V_\mathrm{vir} \right )^{2}}\right],
\label{eq:BH growth quasar}
\end{equation}
where $M_\mathrm{BH,sat}$ is the central black hole mass of the satellite galaxy, $M_\mathrm{sat}$ and $M_\mathrm{cen}$ are the  masses (cold gas and stars) of the satellite and the central galaxy, respectively, $M_\mathrm{cold}$ is the total mass of the cold gas in the satellite and central galaxies, $V_\mathrm{vir}$ is the virial velocity of the central dark matter halo, and $f=0.03$ is a parameter that describes the growth rate of the black hole. Once these black holes reach a sufficient mass (such as $>10^{7} M_{\odot}$), they begin to accrete hot gas from the host galaxy in a steady and continuous manner, releasing an enormous amount of energy into the surrounding circumgalactic medium. This process is known as the 'radio mode' and is associated with the suppression of star formation. Such growth rate via absorbing hot gas is:
\begin{equation}
  \dot{M}_\mathrm{BH}=\kappa \left ( \frac{M_\mathrm{hot}/M_\mathrm{vir}}{0.1} \right )\left ( \frac{V_\mathrm{vir}}{200\mathrm{km}/\mathrm{s}} \right )^{3}\left ( \frac{M_\mathrm{BH}}{10^{8}/hM_{\odot }} \right )M_{\odot }/ yr
\label{eq:BH growth radio}
\end{equation}
where $\kappa$ is a parameter that describes the intensity of absorption, $M_\mathrm{hot}$ is the mass of the hot gas, and $M_\mathrm{vir}$ and $V_\mathrm{vir}$ are the virial mass and velocity of the halo, respectively. While the former mode is the main driver of massive black hole growth in the early stages, the latter dominates at lower redshifts. 

In GABE, the assumption is made that two massive black holes merge instantaneously when their host galaxies merge. This assumption, however, is an oversimplification, as the actual process of two massive black holes dissipating their orbital energy and sinking to the center of the potential well before coalescence involves several physical processes. These processes include dynamical friction with the stellar and gaseous background, three-body interactions with core stars, gravitational and viscous torques of the circumbinary disk, and the emission of GWs when the separation decreases to $\sim0.01\,\mathrm{pc}$ scale. These dissipative processes can span a few billion years \citep{2002MNRAS.331..935Y,2015ApJ...810...49V,2015MNRAS.454L..66S}. Despite this, the disk and bulge configuration of galaxies considered in GABE are too simple to account for these physical processes. Therefore, as did in \cite{2019MNRAS.483..503Y} and \cite{2022A&A...660A..68C},  no time delay between massive black hole binary mergers and galaxy mergers is considered in our work. The Chabrier \citep{2003PASP..115..763C} initial mass function (IMF) is adopted in GABE, while the rapid binary population synthesis model COSMIC v3.3.0 adopts the Kroupa IMF \citep{2001MNRAS.322..231K}. However, the Kroupa IMF and Chabrier IMF are  similar for $M_{\ast} > 1 M_{\odot}$, so we assume no calibration is required in the SGWB calculation especially in Eq. (\ref{eq:merger number density in universe}).

\subsection{Merging stellar compact binaries}
\label{subsec:Stellar mass merging compact binaries}

We have utilized the rapid binary population synthesis model COSMIC v3.3.0 \citep{2020ApJ...898...71B} to derive properties of simple stellar populations (SSPs). The model is based on the binary star evolution code presented in \cite{2002MNRAS.329..897H}, and has been substantially enhanced to incorporate prescriptions for massive star evolution and binary interactions. The primary objective of COSMIC is to simulate compact binaries and their progenitors, making it particularly well-suited to our research goals.

For every SSP with metallicity $Z$, we calculate event numbers of  stellar compact binary mergers per solar mass $\Delta N_{\mathrm{SSP},i}(Z, t_\mathrm{delay}, m_1, m_2)$. Here, $i$ represents categories of merging stellar compact binaries: BNSs, NSBHs, and BBHs; $m_1$ and $ m_2$ denote the masses of the primary and secondary stars, respectively. We divide $m_1$ and $ m_2$ into equal-width bins in the ranges of $1.24M_{\odot}-3 M_{\odot}$ for neutron stars and $3M_{\odot}-45 M_{\odot}$ for black holes.  The delay time $t_\mathrm{delay}$ is the duration between the merger time and the formation time of their progenitor zero age main sequence binaries. We consider the cases of $Z=$ $0.0001$, $0.0003$, $0.001$, $0.004$, $0.01$, $0.02$, and $0.03$ in our work. 

\begin{figure*}
	\includegraphics[width=2.2\columnwidth]{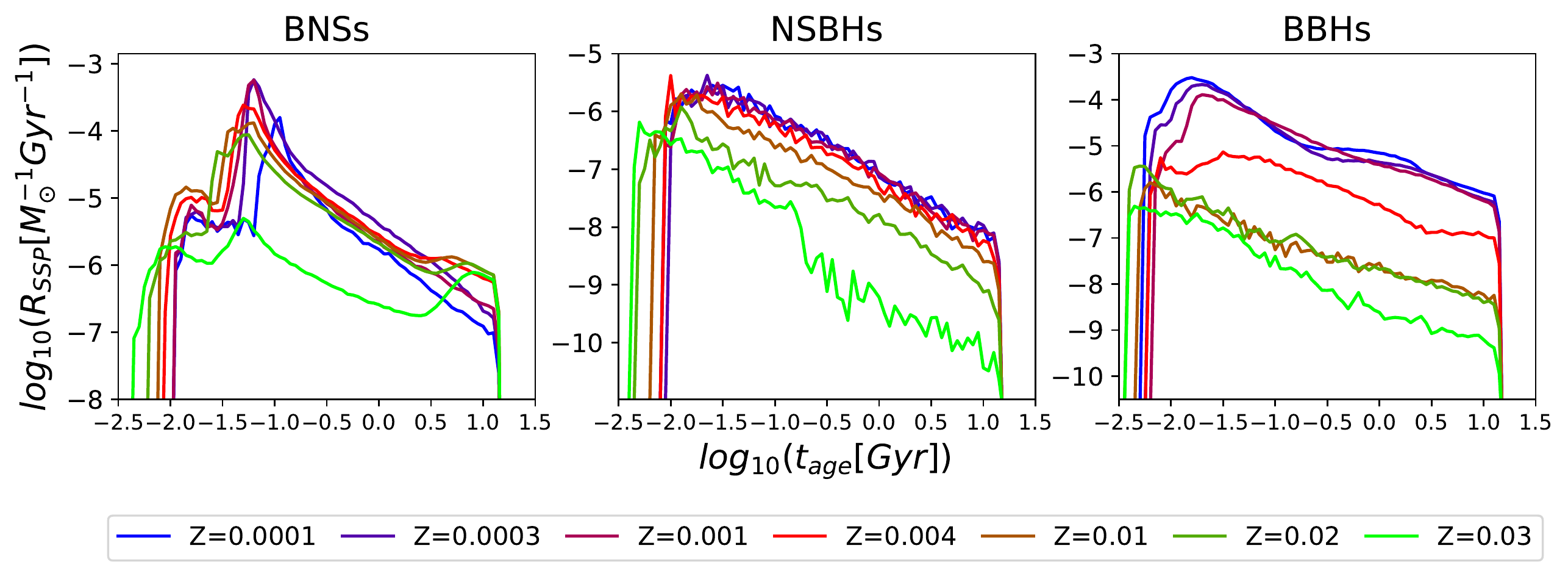}
    \caption{The delay time distribution $R_\text{SSP}$ of merging BNSs, NSBHs, and BBHs is shown from left to right for seven different SSPs with varying metallicities. The age of the SSPs is presented along the $\mathrm{x}$ axis.}
    \label{fig:Delay time distribution}
\end{figure*}

The COSMIC initialization is configured based on the study by \citet{2020ApJ...899L...1Z}, wherein Table 1 lists various options for the initialization, and we adopt the $[$Initial conditions, CE survival, CE efficiency, Remnant mass, Natal kicks$]$ $=[$S$+$2012, Pessimistic, 1.0, Delayed, Bimodal$]$ option for this investigation. Fig.~\ref{fig:Delay time distribution} presents the delay time distributions of the BNSs, NSBHs, and BBHs mergers in the context of SSPs. For all types of merging compact stellar binaries and metallicities, the event rate $R_\mathrm{SSP}$ follows an almost power law distribution $\propto t_\mathrm{delay}^{-1}$ after reaching its peak, as several studies have previously pointed out \citep[e.g.,][]{2012ApJ...759...52D}.

To self-consistently derive the distribution of merging stellar compact binaries in the universe, we apply our SSPs to galaxy catalogues generated by GABE. In GABE, a star formation activity is triggered if the gaseous disk is too massive to be stable or a galaxy merger occurs. Galaxies exhibit star formation events triggered by the disk instability that occurs when the cold gas within their disks surpasses a critical mass \citep{2006MNRAS.365...11C}:
\begin{equation}
    M_\mathrm{crit}=11.5\times 10^{9}\left ( \frac{V_\mathrm{max}}{200\mathrm{km}/\mathrm{s}} \right )\left ( \frac{R_\mathrm{gas,d}}{10\mathrm{kpc}} \right )M_{\odot },
\label{eq:critical mass}
\end{equation}
where $V_\mathrm{{max}}$ represents the maximum circular velocity of the dark matter halo, and $R_\mathrm{{gas,d}}$ denotes the scale radius of the cold gas disk. Subsequently, the cold gas within the disk undergoes conversion into stars at a rate denoted by $\dot{M_{\ast}}$:
\begin{equation}
    \dot{M_{\ast}}=\alpha (M_\mathrm{gas}-M_\mathrm{crit})/t_\mathrm{dyn},
\label{eq:star formation rate disk}
\end{equation}
where $\alpha = 0.02$ is a parameter which describes the star forming efficiency and $t_\mathrm{dyn} = 3R_\mathrm{gas,d}/V_\mathrm{max}$ is the characteristic timescale at
the edge of the star-forming disk. When galaxies undergo mergers, the process can lead to star formation events known as star bursts. During these episodes, the cold gas present in the galaxies undergoes conversion into stars at a specific mass ratio \citep{2001MNRAS.320..504S}:
\begin{equation}
    e_\mathrm{burst} = 0.56(\frac{M_\mathrm{sat}}{M_\mathrm{cen}})^{0.7},
\label{eq:star formation rate merger}
\end{equation}
where $M_\mathrm{sat}$ and $M_\mathrm{cen}$ are the total mass (cold gas mass and star mass) of the satellite and central galaxies, respectively.

After a star formation activity is triggered, the subsequent stellar evolution is assumed to be completed in an instantaneous way in GABE. Stars evolve to their final states (e.g. $t_{\rm age}\sim10{\,{\rm Gyr}}$) right after their birth. During this process, $43\%$ of the stellar mass is recycled back to the circumstance through stellar winds and supernovae, while metals are produced with a yield of $3\%$ of the initial stellar mass. These newly created metals are thoroughly mixed with the pre-existing cold gas, and spread to other galaxy components through baryonic cycle in galaxies. The resulting metal enrichment of galaxies in GABE exhibits a reasonable agreement with observational mass-metallicity relations \citep[see Fig.~2 and section~2.2.7 of][]{2019RAA....19..151J}. Note that the instantaneous assumption is only used to calculate the recycled mass and metals. The detailed evolution of compact binaries and the time delay of each merger have been specifically considered, as described below.

Each star formation event is treated as the birth of a SSP and the corresponding information of the SSP, such as the formation time, mass, and metallicity, is recorded. This method allows us to generate star formation histories for galaxies of various types.
Subsequently, the total number of mergers within the simulation volume in the redshift interval $\Delta z$ is calculated  by:
\begin{multline}
\Delta N_i(z, m_1, m_2) = \\
\\ \sum_{k=0}^{n_\mathrm{SSP}} M_k \sum_{a=0}^{a_\mathrm{max}}\Delta N_{\mathrm{SSP},ika}\left ( Z_k, t_{\mathrm{delay},ika}(z), m_1, m_2 \right ),
\label{eq:merger number density in universe}
\end{multline}
where $i$ represents species of merging stellar compact binaries: BNSs, NSBHs or BBHs; $M_k$ and $Z_k$ are mass and metallicity of the $k$-th SSP respectively, determined self-consistently in GABE; $n_\mathrm{SSP}$ is the total number of the SSPs; $t_{\mathrm{delay},ika}$ is $a$-th delay time bin of kind $i$ object of the $k$-th SSP; and $a_\mathrm{max}$ is the number of delay time bins in $k$-th SSP needed to cover the redshift interval $\Delta z$. The equation reveals that $\Delta N_i$ in the redshift interval $\Delta z$ is contributed by the stellar compact binaries that will merge in this interval (the secondary sum over $a$) from all SSPs (the first sum over $k$). See more details in  Appendix. Thus, by utilizing the aforementioned approach, we are able to obtain the numbers and rates of merging events for various types of stellar compact binaries. The findings from this analysis are subsequently presented and discussed in the following section.

It is worth noting that, according to the binary population synthesis model COSMIC v3.3.0, the mass of the heaviest stellar black holes is estimated to be around $45 M_{\odot}$, while the lightest massive black holes have a mass of approximately $100 M_{\odot}$ with the light-seed model (MSII). It is important to emphasize that no channels have been introduced in this work to generate black holes within the pair-instability mass gap. Nevertheless, merging BBHs have been observed within this mass gap, such as the case of GW190521 \citep{2020PhRvL.125j1102A}, which may involve a primary black hole of $85 M_{\odot}$ and a secondary black hole of $66 M_{\odot}$. It is suggested that events similar to GW190521 may have a dynamical origin \citep{2020ApJ...903L...5R}, or even a cosmological origin \citep{2020arXiv200706481C}, but this goes beyond the scope of our current investigation.

\section{Results}
\label{sec:Results}

\subsection{Event numbers and merger rates}

\begin{figure*}
    \includegraphics[width=1.33\columnwidth]{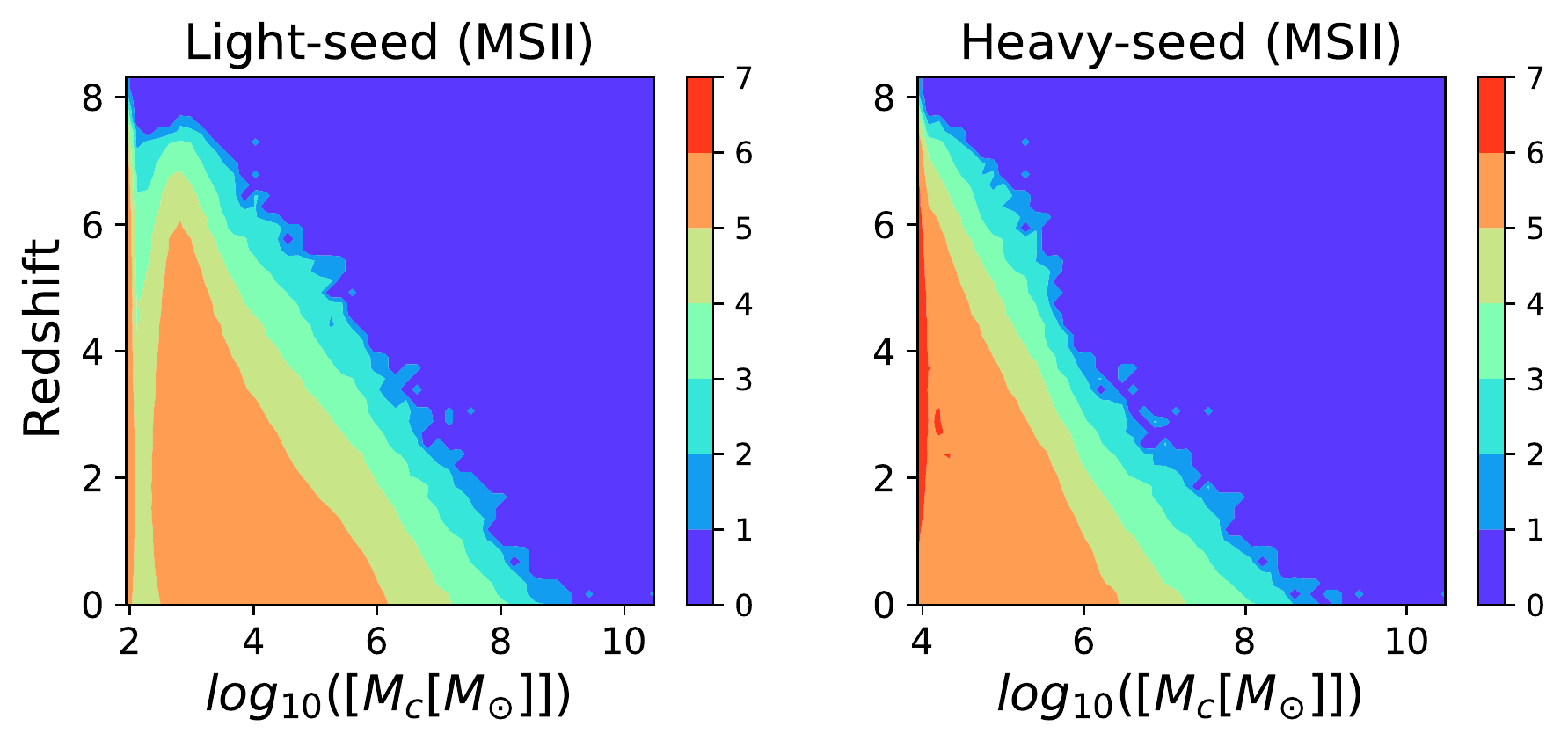}
    \includegraphics[width=2\columnwidth]{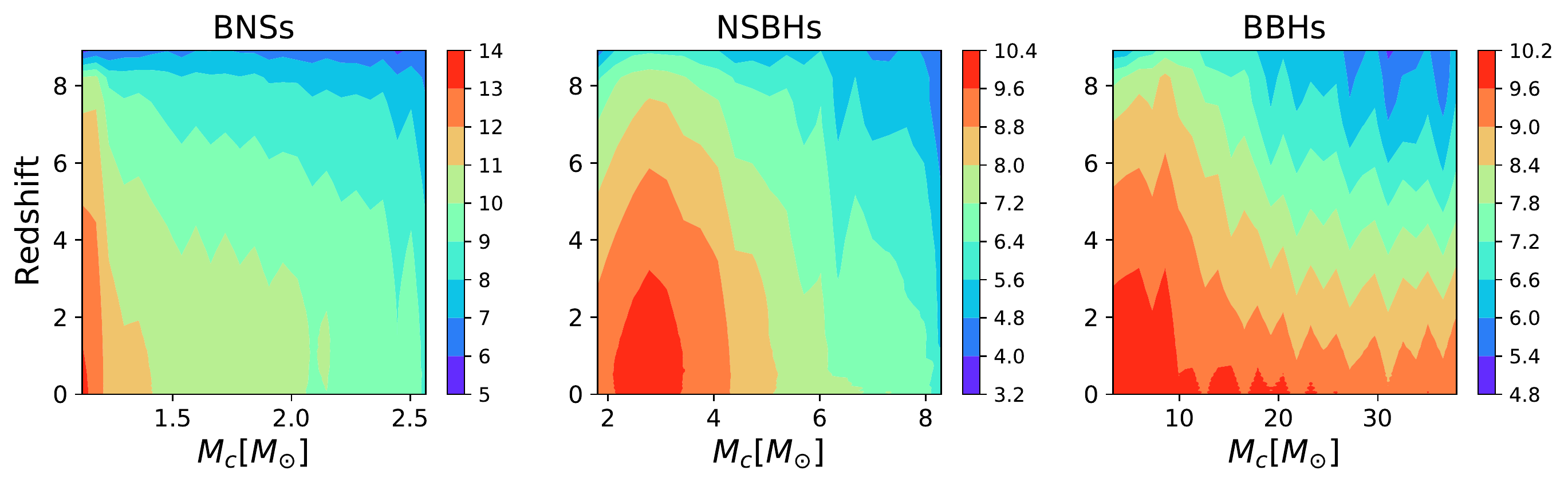}
    \caption{Upper panel: the event number distribution $N$ of merging massive black hole binaries over chirp mass and redshift, computed using the MSII simulation in two scenarios of light-seed and heavy-seed models. The values of event numbers are indicated by color bars in the unit of $\text{log}_{10}(\text{d}^2 N/(\text{d} z \text{d} \text{log}_{10}M_c))$. Lower panel: the event number $N$ distribution of merging stellar compact binaries over chirp mass and redshift. The values of event numbers are indicated by color bars in the unit of $\text{log}_{10}(\text{d}^2 N/(\text{d} z  \text{d} M_c))$.}
    \label{fig:massive black hole event rates}
\end{figure*}

\begin{figure}
	\centering
	\includegraphics[width=\columnwidth]{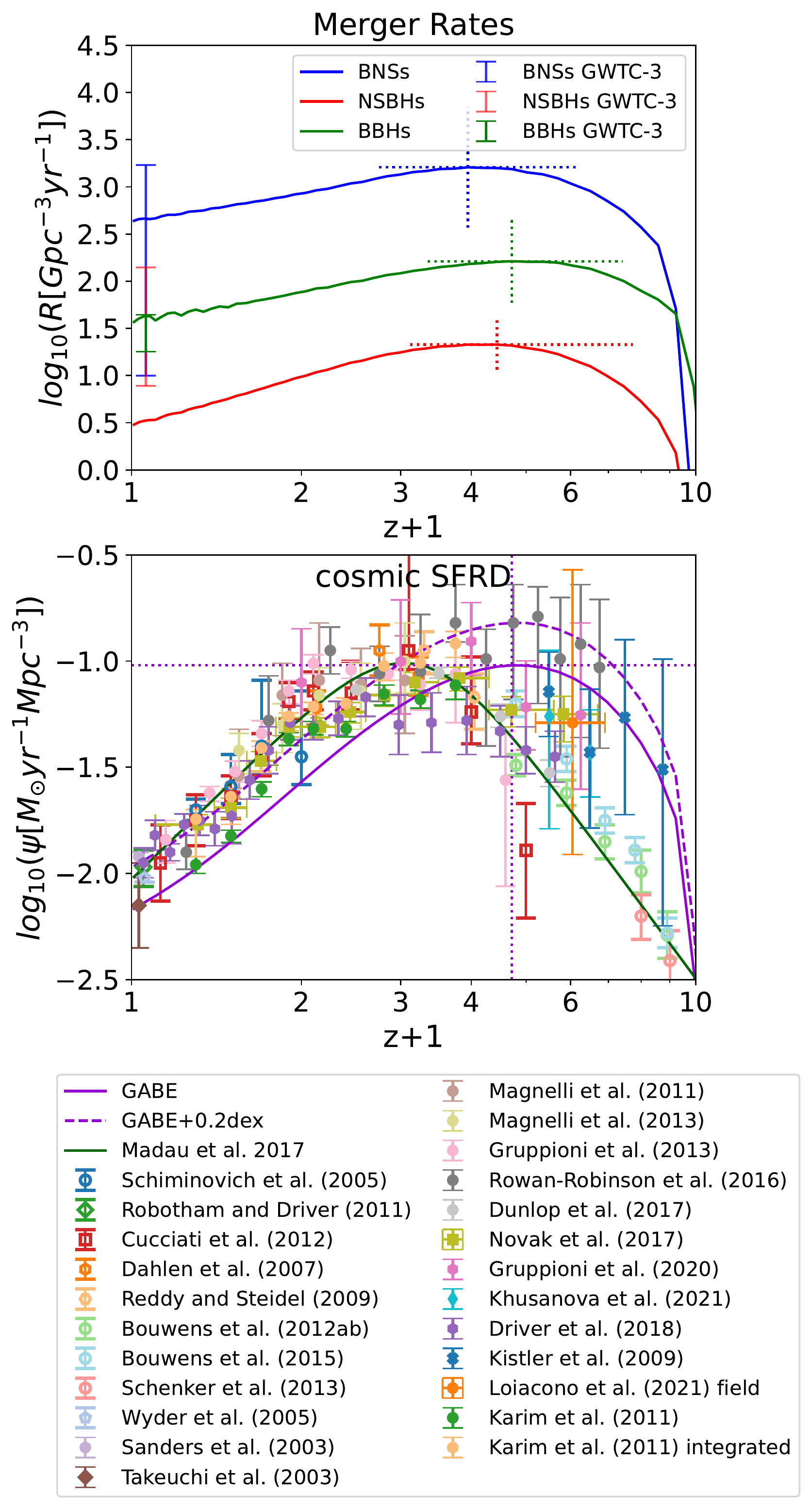}
	\quad
    \caption{Upper panel: The merger rates of BNSs, NSBHs and BBHs in the universe calculated from GABE. The blue, red and green bars are the estimation range of local merger rates \citep{2021arXiv211103634T} according to GWTC-3. Lower panel: The SFRD in GABE, together with the SFRD in \citet{2017ApJ...840...39M}. The empirical constraints for the SFRD available from the literature based on multiwavelength observations are also shown. All the data points and SFRD values correspond to a Chabrier IMF \citep{2003PASP..115..763C}. The peaks of the merger rates and GABE-derived SFRD are marked by the intersection of two dotted lines.}
    \label{fig:cSFRs-rate}
\end{figure}

We present the event number distributions of merging massive black hole binaries in the MSII simulation over chirp mass and redshift, where both light-seed and heavy-seed models are taken into account. The resulting distributions are illustrated in the upper panels of Fig.~\ref{fig:massive black hole event rates}. Our analysis reveals that the redshift-chirp mass distribution of both seed models is quite similar above the chirp mass $M_c=10^6 M_\odot$, whereas it becomes increasingly distinct with decreasing chirp mass. Additionally, we also display the event number distributions of merging stellar compact binaries, which are found to be independent of the seed models, in the lower panels of the same figure. In the lower right panel of Fig.~\ref{fig:massive black hole event rates}, we notice that most massive BBH mergers occur at low redshift ($z<2$). These merging BBHs of high mass actually form in metal-poor SSPs ($Z\lesssim0.004$) at high redshift, while they exhibit significant delay times, ranging from $\sim1\Gyr$ to the age of the universe. As a result, these high-mass BBHs do not undergo immediate merging after their birth; instead, their mergers occur at low redshifts.

The upper panel of Fig.~\ref{fig:cSFRs-rate} presents the merger rates of different types of stellar binaries. Specifically, the local merger rates of BNSs and BBHs are consistent with those reported in the third Gravitational-Wave Transient Catalog (GWTC-3) \citep{2021arXiv211103634T}, at $436.8$ and $37.0$ $\mathrm{Gpc}^{-3}\mathrm{yr}^{-1}$, respectively. However, the local merger rate of NSBHs is found to be lower, at $3.0 \mathrm{Gpc}^{-3}\mathrm{yr}^{-1}$, compared to that reported in GWTC-3. Possible sources of this discrepancy include uncertainties in the semi-analytic galaxy formation model and binary population synthesis model, as well as the formation channels of binary compact objects. In particular, the event rate may be influenced by the uncertainties in the SFRD and metallicity evolution history in the semi-analytic models, and by the unclear physical processes of binary evolution in the binary population synthesis model. Additionally, our calculation only considers isolated field binaries, whereas merging stellar compact binaries may also form through other channels, such as dynamical encounters and cosmological processes \citep[e.g.][]{2022A&A...660A..26B}.

The redshift evolution of merger rates of BBHs, BNSs, and NSBHs above shows very broad, flat peaks around $z\sim3.7$, $z\sim2.9$, and $z\sim3.4$, spanning a significant redshift range, respectively. These peaks are located at higher redshifts compared to most other studies \citep[e.g., Fig.~6 in][]{2022MNRAS.516.3297S}, primarily because the SFRD derived by GABE is slightly different from that in other studies. We display the SFRD from GABE and compare it with that from \citet{2017ApJ...840...39M}, in the lower panel of Fig.~\ref{fig:cSFRs-rate}. We also present various observational estimates compiled by \citet{2021A&A...646A..76L}, including, the results from UV surveys \citep{2005ApJ...619L..47S, 2005ApJ...619L..15W, 2007ApJ...654..172D, 2009ApJ...692..778R, 2011MNRAS.413.2570R, 2012ApJ...752L...5B, 2012ApJ...754...83B, 2012A&A...539A..31C, 2013ApJ...768..196S, 2015ApJ...803...34B}, the results from infrared, mm, and radio-selected galaxies \citep{2003AJ....126.1607S, 2003ApJ...587L..89T, 2011ApJ...730...61K,2011A&A...528A..35M, 2013A&A...553A.132M, 2013MNRAS.432...23G, 2016MNRAS.461.1100R, 2017MNRAS.466..861D, 2017A&A...602A...5N}, the result from an optical–NIR observation \citep{2018MNRAS.475.2891D}, and the result from gamma–ray bursts \citep{2009ApJ...705L.104K}. In addition, we include estimates from the ALPINE collaboration \citep{2020A&A...643A...8G, 2021A&A...646A..76L, 2021A&A...649A.152K}. Below z $\sim 2$, the SFRD from GABE is lower than the median of the observational results by $\sim 0.2$ dex (see the dashed line plotted in the lower panel of Fig.~\ref{fig:cSFRs-rate}, which is boosted by this factor.). At higher z, there are large scatters in the observational results of the SFRD, the SFRD from GABE is consistent with some observations \citep[e.g.,][]{2009ApJ...705L.104K, 2016MNRAS.461.1100R, 2020A&A...643A...8G, 2021A&A...646A..76L}. The noteworthy feature of the SFRD from GABE is its peak at higher redshifts (z$\sim4$), which distinguishes it from the SFRD presented in \citet{2017ApJ...840...39M}. Notably, some other semi-analytic models also exhibit broad peaks, albeit occurring at lower redshifts \citep[e.g. see Fig.~11 of][]{2015MNRAS.451.2663H}.

The lower SFRD at $z<2$ in our model certainly underestimates the merger rates of the stellar compact binaries in the redshift range by about $0.2$ dex at most. This upper limit is derived assuming an extreme case where the SFRD is underestimated throughout all redshift ranges. We should keep this as a caveat to the results relevant to the stellar compact binaries in the following sections.

\begin{figure}
	\centering
	\includegraphics[width=\columnwidth]{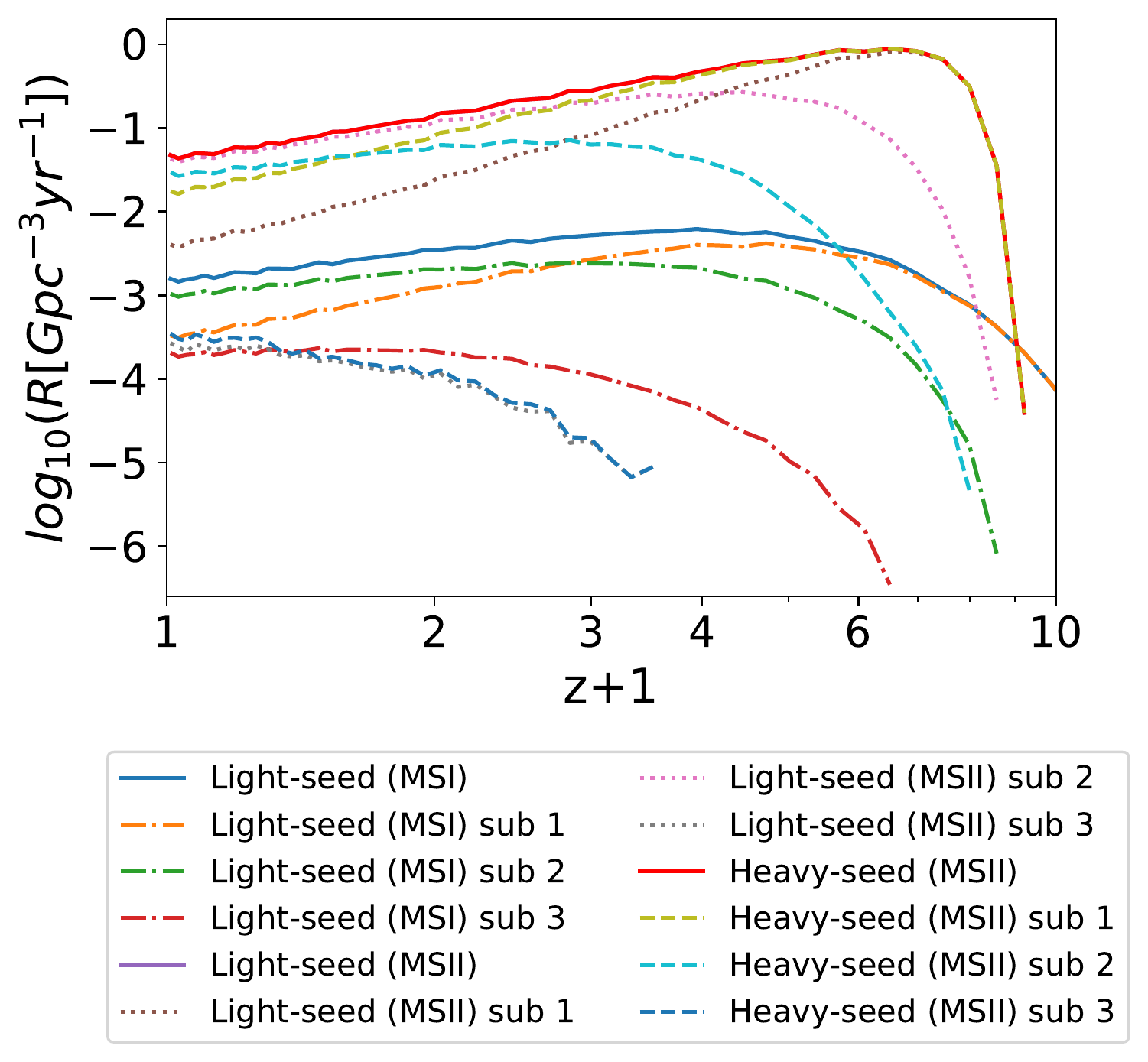}
	\quad
    \caption{The merger rates of massive black hole binaries in the universe calculated using GABE. The total merger rates of the MSI and MSII models are shown in solid lines. Three sub-populations are divided based on the chirp mass $M_c$ of these binaries: sub1, $M_c<8.7\times10^2M_\odot$ for light-seed and $M_c<8.7\times10^4M_\odot$ for heavy-seed model; sub2, $8.7\times10^2 \leq M_c<8.7\times10^6M_\odot$ for light-seed and $8.7\times10^4 \leq M_c<8.7\times10^6M_\odot$ for heavy-seed; and sub3, $M_c\geq8.7\times10^6M_\odot$}
    \label{fig:SMBHrates}
\end{figure}

Fig.~\ref{fig:SMBHrates} shows the merger rates of massive black hole binaries, including the light-seed (MSI) model and both the light- and heavy-seed (MSII) models, represented by solid lines in different colors. The MSI model exhibits approximately $1.5-2$ magnitudes lower merger rates than the MSII models, attributed to its 125 times worse particle mass resolution. Note that the two solid lines of the MSII models are exactly the same and overlap with each other, as they actually share the same dark matter halo merger trees and only the embedded seed mass is different. We also divide the mergers into different sub-populations according to their chirp masses: sub1 ($M_c<8.7\times10^2M_\odot$ for light-seed and $M_c<8.7\times10^4M_\odot$ for heavy-seed); sub2 ($8.7\times10^2 \leq M_c<8.7\times10^6M_\odot$ for light-seed and $8.7\times10^4 \leq M_c<8.7\times10^6M_\odot$ for heavy-seed); and sub3 ($M_c\geq8.7\times10^6M_\odot$). Agreed with the hierarchical clustering of $\Lambda$CDM cosmology, massive black hole binary mergers are mostly contributed by sub1 at the beginning and are then generally dominated by sub2. Though sub3 has the highest chirp mass, its event rate is still the lowest even at $z\sim0$.

\subsection{The multi-band SGWB}\label{sec:multi-sgwb}

\begin{figure*}
	\includegraphics[width=2\columnwidth]{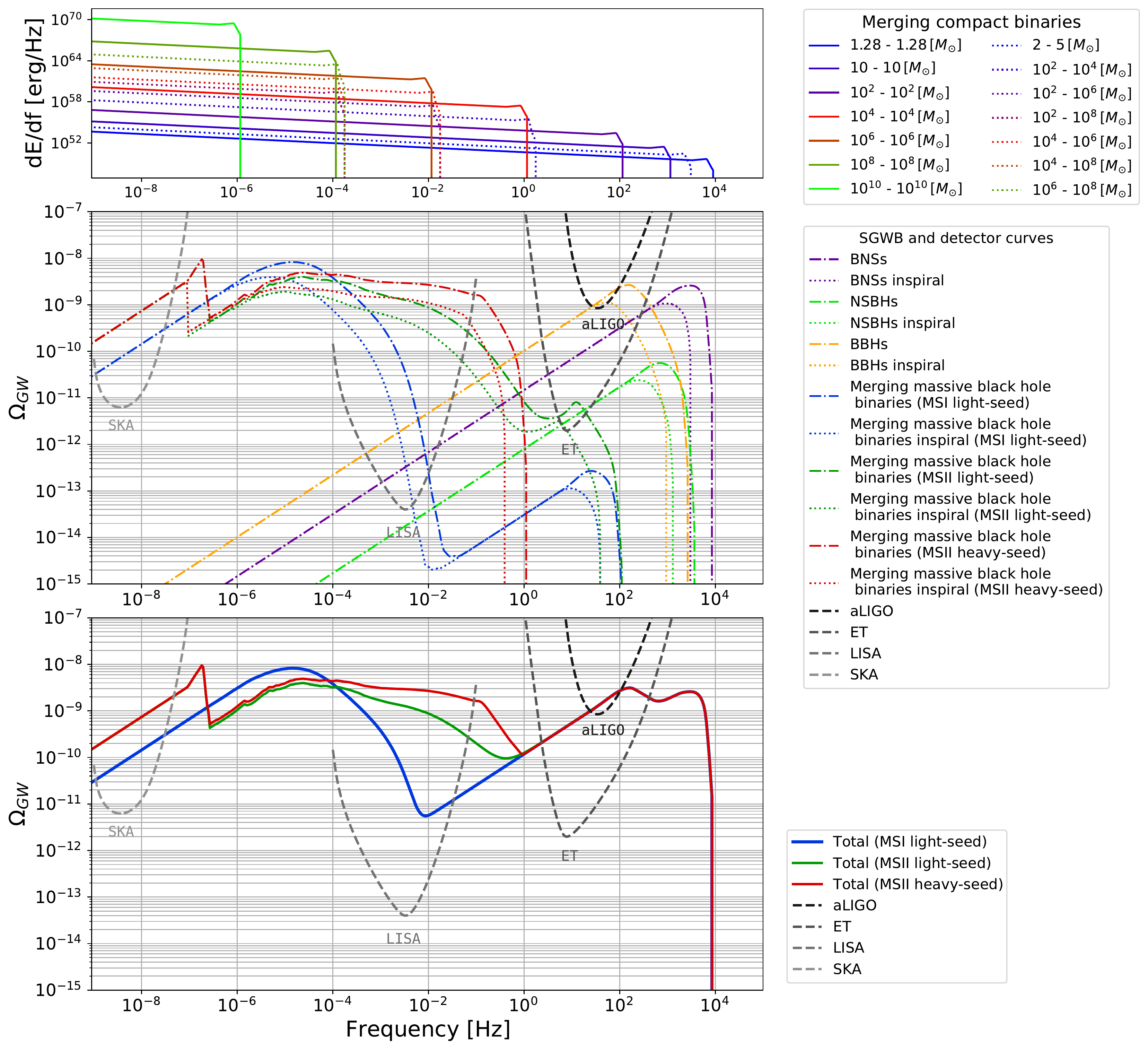}
    \caption{Upper panel: the GW energy spectra of various merging systems, including BNSs, NSBHs, BBHs, and massive black hole binaries. The spectra are calculated using Eq.~(\ref{eq:gw wave}). Middle panel: the SGWB from merging massive black hole binaries with different seed models, as well as merging BNSs, NSBHs, and BBHs. The inspiral phases are plotted separately. Lower panel: the total multi-band SGWB. In addition, the sensitivity curves of aLIGO, ET, LISA, and SKA are shown in both the middle and lower panels.}
    \label{fig:SGWB total}
\end{figure*}

By utilizing the merger rates and GW energy spectra, the SGWB contributed by massive black hole binaries and stellar binaries can be obtained via Eqs.~(\ref{eq:Omega in use directly}) and (\ref{eq:Omega in use}), respectively. The middle panel of Fig.\ref{fig:SGWB total} illustrates the SGWB spectra produced from different sources. Moreover, the total SGWB generated from all considered sources is presented in the lower panel of Fig.\ref{fig:SGWB total} in the frequency range spanning from nHz to kHz. To investigate the influence of the seed black hole models, we calculate the SGWB with two different seed models. Note that we only present the results from the MSI with the light-seed model, because we only use ultra-low frequency band of the MSI results which are unaffected by seed models. In general, the amplitudes of the principal peaks of the SGWB energy density are within one order of magnitude in all two MSII cases under consideration, i.e., $\Omega_{\mathrm{GW}} \sim 10^{-9}$ to $10^{-8}$.
Furthermore, to illustrate the impact of GW energy spectra on SGWB spectra, we calculate the GW energy spectra for binaries with several typical masses using Eq.(\ref{eq:gw wave}), and present them in the upper panel of Fig.~\ref{fig:SGWB total}.

To predict the detectability of the SGWB, the PI sensitivity curves of current and forthcoming detectors are plotted as guidelines in the middle and lower panels of Fig. \ref{fig:SGWB total}. The threshold SNR is taken as $\mathrm{SNR}_{\mathrm{th}}=3$, and the observation periods for ground-based detectors (i.e., aLIGO and ET) and LISA are assumed to be $T = 1$ year and $T = 4$ years, respectively. For the SKA, we adopt the number of pulsars as $N_\mathrm{p} = 200$, the rms timing residual $\sigma = 50$ ns with no red noise, the observing period of $T = 10$ years, and the average observation cadence of $1$ per week \citep{2021JCAP...01..012C, 2019arXiv191109745M}. 

In the ultra-low frequency band ($\sim10^{-9}-10^{-7} \mathrm{Hz}$) which are mainly contributed by merging supermassive black hole binaries, the SGWB amplitude predicted by the MSII is almost one order of magnitude higher than that of the MSI. This is simply because the inadequate volume of the MSII , which results in poor statistics for extremely large supermassive black holes, leading to noisy predictions in this frequency band (see section \ref{subsec:Massive merging compact binaries} for more details). We report the dimensionless characteristic strain $A_{\mathrm{yr}^{-1}}$ as $5.1 \times 10^{-16}$, which falls within the same order of magnitude as those predicted by previous theoretical studies and observed by previous experiments\footnote{For instance, theoretical predictions include $A_{\mathrm{yr}^{-1}}=4 \times 10^{-16}$ \citep{2016MNRAS.463L...6S}, $7.1 \times 10^{-16}$ \citep{2017MNRAS.464.3131K}, $5 \times 10^{-16}$ \citep{2019MNRAS.483..503Y}, $< 2.4 \times 10^{-15}$ \citep{{2019MNRAS.482.2588Z}}, $1.4 \times 10^{-16}$ \citep{2022A&A...660A..68C}, $1.2 \times 10^{-15}$ \citep{2022MNRAS.509.3488I}, and $6.4 \times 10^{-16}$ \citep{2022MNRAS.511.5241S}. The observations from EPTA, NANOGrav, PPTA, and IPTA establish the upper limits of $A_{\mathrm{yr}^{-1}}$ as $3\times 10^{-15}$ \citep{2015MNRAS.453.2576L}, $1.45 \times 10^{-15}$ \citep{2018ApJ...859...47A}, $1\times 10^{-15}$ \citep{2015Sci...349.1522S}, and $1.7 \times 10^{-15}$ \citep{2016MNRAS.458.1267V}, respectively. But see recent results by EPTA, NANOGrav, PPTA, and CPTA: $A_{\mathrm{yr}^{-1}}= 2.5^{+0.7}_{-0.7} \times 10^{-15}$ \citep{2023arXiv230616214A}, $2.4^{+0.7}_{-0.6} \times 10^{-15}$ \citep{2023ApJ...951L...8A}, $2.04^{+0.25}_{-0.22} \times 10^{-15}$ \citep{2023ApJ...951L...6R} and $\text{log}A_c = -14.7^{+0.9}_{-1.9}$ \citep{2023RAA....23g5024X}, respectively.} The sensitivity curve of SKA shows its capability of detecting the SGWB across the ultra-low frequency range from $\sim 10^{-9}$ to $10^{-7} \mathrm{Hz}$, following the typical power law $\Omega_{GW} \propto f^{2/3}$.

In the low frequency band ($\sim10^{-4}-1 \mathrm{Hz}$), the SGWB spectra obtained from the MSII have much flatter shapes and higher amplitudes compared to those of the MSI. These differences simply reflect the fact that a larger number of intermediate-mass massive black holes are not resolved in the MSI. These suggest that the $\Omega _\mathrm{GW}$ signal can be detected by LISA in this frequency band, irrespective of the seed model employed. Furthermore, the middle panel of Fig. \ref{fig:SGWB total} indicates that the merging massive black hole binaries contributes substantially to the SGWB in this frequency range. Within the frequency range of approximately $10^{-4}$ to $10^{-2}$ Hz, LISA has the ability to detect the SGWB generated by merging BBHs and BNSs. Interestingly, the SGWB signal predicted by the heavy-seed model exhibits a higher and flatter spectrum than that of the light-seed model in the same frequency band. This is due to the higher event number of massive black holes in the range of $\sim 10^4 - 10^6 M_{\odot}$ in the heavy-seed model, as demonstrated in the upper panel of Fig.~\ref{fig:massive black hole event rates}. These sources dominate the GW energy spectra in the LISA band, as shown in the upper panel of Fig. \ref{fig:SGWB total}, hence enabling LISA to potentially discriminate between black hole seed models. 

In the high frequency band ($\sim1- 10^{4} \mathrm{Hz}$), the SGWB is primarily dominated by the contribution from merging BBHs ( $\sim 1$ to $100 \mathrm{Hz}$) and  merging BNSs ($\sim 10^3$ to $10^4 \mathrm{Hz}$). Specifically, at $\sim 100 \mathrm{Hz}$, the $\Omega _\mathrm{GW}$ contribution from merging BBHs is $\sim 10$ and $\sim 100$ times larger than that from merging BNSs and NSBHs, respectively. It is worth noting that the contribution from merging NSBHs is negligible, primarily due to their low event rates, as shown in Fig.~\ref{fig:cSFRs-rate}. In the frequency range of $\sim 100$ to $10^3 \mathrm{Hz}$, the total SGWB spectra exhibit small variations, with amplitudes of $\Omega_{\mathrm{GW}} \sim 2\times 10^{-9}$. It should be noted that this outcome is dependent on the adopted GW energy spectrum of BNSs presented in Eq.~(\ref{eq:gw wave}). Typically, the energy spectra of BNSs are truncated at the frequency of innermost stable circular orbit ($\sim 800\mathrm{Hz}$) for simplicity, and the post-merger phases of BNSs are not considered \citep[e.g.][]{2018PhRvL.120i1101A}. As depicted in the middle panel of Fig. \ref{fig:SGWB total}, in the frequency range of ground-based detectors, the SGWB spectra are primarily dominated by the inspiral stages of the binary systems, consistent with the findings of previous research \citep[e.g.][]{2016PhRvL.116m1102A, 2017PhRvL.118l1101A, 2011PhRvD..84l4037M}. In addition, the ET has the capability to detect the SGWB emanating from not only stellar binaries but also massive black hole binaries. The $\Omega_{GW}$ from merging massive black hole binaries in the light seed scenario displays double peaks. The peaks at $\sim10$ Hz are mostly caused by the first generation of mergers involving massive black holes with masses close to their initial seed mass (e.g. $\le2$ times the light seed mass). These massive black holes do not undergo a substantial accretion phase. The SGWB spectra obtained in this study exhibit similar shapes to those from previous works, as reported in \citep[e.g.,][]{2019ApJ...871...97C,2021JCAP...11..032C}. We also note, similar to the merger rates in our model, the estimation of the SGWB could also be underestimated by $\sim 0.2$ dex at most.

\subsection{Detection ability}
\begin{table}
	\centering
	\caption{The SNR for aLIGO, ET, LISA and SKA with light-seed (MSI), light-seed (MSII) and heavy-seed (MSII) model. The observing periods  are assumed as $1$, $4$ and $10$ year(s) for ground-based detectors, LISA and SKA, respectively. For SKA we adopt $N_\mathrm{p} = 200$ pulsars, the rms timing residual $\sigma = 50$ ns with no red noise, and the average observation cadence of $1$ per week.} 
	\label{tab:SNR}
	\begin{tabular}{lcccr} 
		\hline
		 & Light-seed (MSI) & Light-seed (MSII) & Heavy-seed (MSII)\\
		\hline
		aLIGO (1yr) & / & 5.01 & 5.00\\ 
        ET (1yr) & / & 744.60 & 736.73\\
        LISA (4yrs) & / & 67159.63 & 155885.39\\
        SKA (10yrs) & 42.84 & / & /\\
		\hline
	\end{tabular}
\end{table}

In addition to estimating from PI sensitivity curves, we have calculated the SNR of the SGWB, as defined in Section \ref{subsec:Signal to Noise ratio}, to evaluate the detectability of upcoming detectors. We present the results in Table~\ref{tab:SNR}. The obtained SNRs for aLIGO, ET, and SKA in the given observing periods are approximately $5$, $700$, and $42.84$, respectively (note that the obtained SNRs for aLIGO and ET might be underestimated up to $\sim 0.2$ dex. In any case, the SNR is much higher than a detection SNR threshold. ). However, the SNR from LISA exhibits significant dependence on the adopted black hole seed model, yielding values of approximately $6.7 \times 10^4$ and $1.5 \times 10^5$ for the light-seed and heavy-seed models, respectively. This indicates that LISA has the potential to differentiate between black hole seed models.

\begin{table}
	\centering
	\caption{The observing time required for aLIGO, ET, LISA, and SKA to achieve a threshold SNR of $3$, with consideration of the light-seed (MSI), light-seed (MSII), and heavy-seed (MSII) models.}
	\label{tab:Time for each experiment needed to get  SNR$=3$.}
	\begin{tabular}{lcccr} 
		\hline
		 & Light-seed (MSI) & Light-seed (MSII) & Heavy-seed (MSII)\\
		\hline
        aLIGO & / & 0.36yr & 0.36yr\\
        ET & / & 513s & 524s\\
        LISA & /  &$\frac{1}{f_{\text{min}}}$& $\frac{1}{f_{\text{min}}}$\\
        SKA & 2.09yr & / & /\\
		\hline
	\end{tabular}
\end{table}
 
It is of significance to estimate the time required for detectors to collect sufficient data and announce detections, which is equivalent to determining the observing periods necessary for a detector to exceed a given threshold SNR. In this study, we have evaluated the observing periods required for the detectors under consideration by setting the threshold SNR as $\mathrm{SNR}_\mathrm{th}=3$. Our findings are presented in Table~\ref{tab:Time for each experiment needed to get SNR$=3$.}. The results indicate that the SGWB is expected to be detected by aLIGO in a relatively short timescale of a few months, while SKA may require a duration of approximately two years. Notably, regardless of the adopted black hole seed model, the ET and LISA detectors may detect the SGWB immediately after the collection of a sufficient duration of data to calculate the SNR. As a result of the underestimated SGWB in the high frequency band in our model, the above estimation of time for detection by aLIGO and ET is conservative.

\section{Summary and Discussion}
\label{sec:Summary and Discussion}

In this study, we explore the potential for the joint detection of the multi-band SGWB from astrophysical sources by future ground-based, space-based detectors and pulsar timing arrays. To this end, we have developed a self-consistent methodology that combines the MSI and MSII simulations with a semi-analytic galaxy formation model--GABE and a binary population synthesis model (COSMIC v3.3.0) to model galaxy and binary evolution in the universe. By treating binary and galaxy evolution in a self-consistent manner, we obtain a comprehensive model of the population of stellar compact binaries (BNSs, NSBHs, BBHs) and massive black hole binaries in the universe. 

Our results show that the total multi-band SGWB from merging massive black hole binaries and stellar compact binaries has main peaks with amplitudes of $\sim$ a few $10^{-9}$ (Fig. \ref{fig:SGWB total}), which could be easily detected by future GW detection experiments at different sensitivity bands (e.g. ET, LISA  and SKA). The shape of the SGWB spectrum in the low frequency band may be influenced by the assumed seed black hole models, which could be determined by LISA. Furthermore, in the high frequency band, the SGWB from merging BBHs is $\sim 10$ and  $\sim 100$ times higher than the SGWB from merging BNSs and merging NSBHs at $\sim 100$ Hz. The SGWB is nearly flat in the $\sim 100-1000$ Hz frequency range, with an amplitude of $\sim 2 \times 10^9$. This result strongly depends on the GW energy spectrum adopted when the frequency is over $\sim 800 $ Hz. We note that the GW energy spectra of BNSs are closely related to the post-merger physics of neutron stars, and thus the SGWB over $\sim 800$ Hz may have the potential to test the equation of state of BNSs.

Note that the SFRD derived from GABE is slightly lower than the observational results by $\sim 0.2$ dex at low redshifts (z$< \sim2$). Consequently, our prediction of the SGWB produced by stellar compact binaries might be underestimated by a similar factor at most. Given uncertainties in other effects, like the time-delay function, metallicity in the binary population synthesis model, our results on high frequency band ($\sim$ [1 , $10^{4}] $ Hz) of the SGWB are still conservative.

\section*{Acknowledgements}
We acknowledge valuable input from our anonymous
referee. This work has been supported by the National Natural Science Foundation of China (Nos.11988101, 11922303, 12033008 and  11673031),  the K. C. Wong Education Foundation,  the National Key Research and Development Program of China (Grant No.2020YFC2201400, SQ2021YFC220045-03) and the Fundamental Research Funds for the Central Universities (No.2042022kf1182). X. Fan. is supported by Hubei province Natural Science Fund for the Distinguished Young Scholars (2019CFA052).

\section*{Data Availability}

This theoretical study did not generate any new data.




\bibliographystyle{mnras}
\bibliography{main.bbl} 

\appendix

\section{The total number of mergers} 
The total number of mergers in the redshift interval is a key ingredient in combining the results of GABE and COSMIC v3.3.0. Here, we illustrate the calculation in detail through a schematic illustration.

\begin{figure*}
	\includegraphics[width=2\columnwidth]{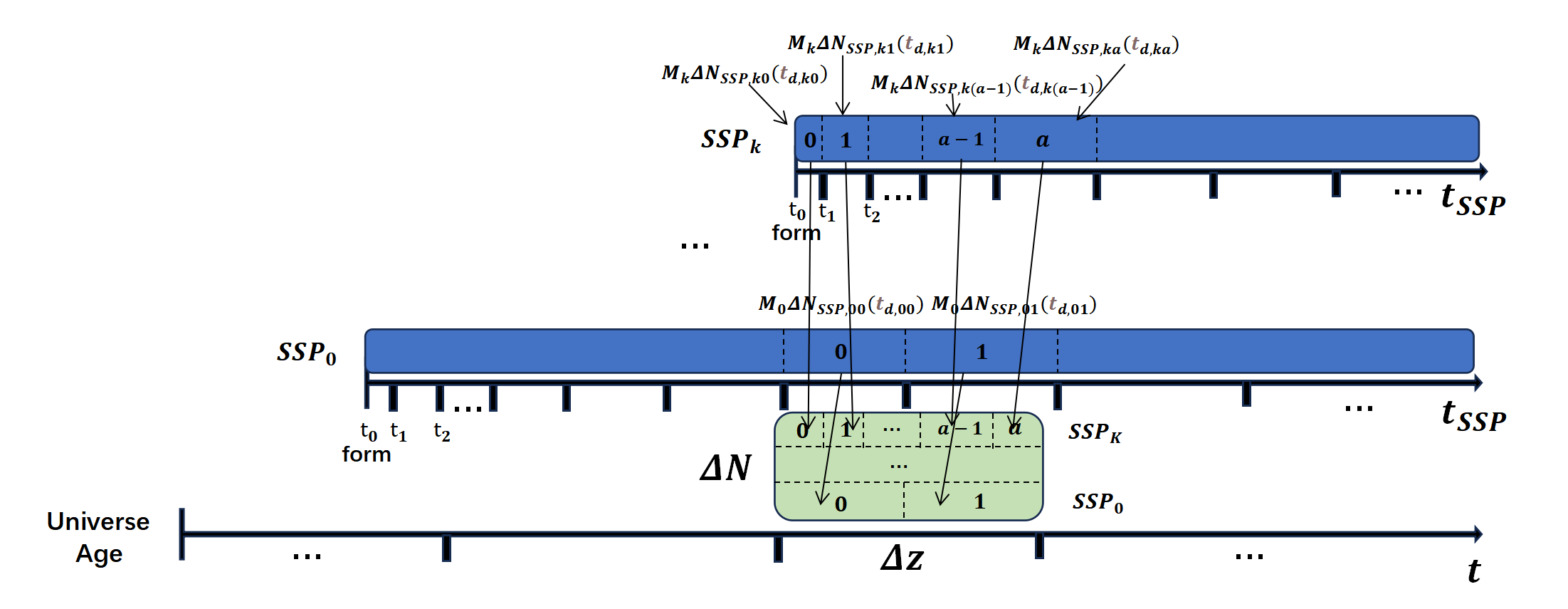}
    \caption{Schematic illustration of Eq.~(\ref{eq:merger number density in universe}). The two arrays shown in blue represent two SSPs formed at $t_0$ respectively, while the green array represents $\Delta N_i(z)$ within a redshift interval $\Delta z$. The thick lines and scales below each array depict the flow of cosmological time.}
    \label{Schematic illustration}
\end{figure*}

The total number of mergers within the simulation volume in the redshift interval $\Delta z$ is calculated  by:
\begin{equation}
\begin{aligned}
\Delta N_i(z, m_1, m_2) &= \\
& \sum_{k=0}^{n_\mathrm{SSP}} M_k \sum_{a=0}^{a_\mathrm{max}}\Delta N_{\mathrm{SSP},ika}\left ( Z_k, t_{\mathrm{delay},ika}(z), m_1, m_2 \right ) \nonumber
\end{aligned}
\end{equation}
where $i$ represents species of merging stellar compact binaries: BNSs, NSBHs or BBHs; $M_k$ and $Z_k$ are mass and metallicity of the $k$-th SSP respectively, determined self-consistently in GABE; $n_\mathrm{SSP}$ is the total number of the SSPs; $t_{\mathrm{delay},ika}$ is $a$-th delay time bin of kind $i$ object of the $k$-th SSP; and $a_\mathrm{max}$ is the number of delay time bins needed to cover the delay time interval of $k$-th SSP. We have created a schematic illustration of Eq.~(\ref{eq:merger number density in universe}) as depicted in Fig. \ref{Schematic illustration}, which shows the relationship between the event number in the universe within the redshift interval $\Delta z$, denoted as $\Delta N$, and the event number per solar mass within the SSP, denoted as $\Delta N_{\text{SSP}}$.

\bsp	
\label{lastpage}
\end{document}